\documentclass[10pt,twocolumn,showpacs,preprintnumbers,amsmath,amssymb,aps,prl,longbibliography,superscriptaddress,footinbib]{revtex4-2}

\usepackage{array,braket,mathtools,siunitx}
\usepackage{bm}

\usepackage[unicode=true,
bookmarks=true,bookmarksnumbered=true,bookmarksopen=false,
breaklinks=true,pdfborder={0 0 0},backref=false,colorlinks=true]{hyperref}
\hypersetup{citecolor={blue},urlcolor={magenta}}

\usepackage{cleveref}
\crefname{appendix}{App.}{Apps.}
\crefname{equation}{Eq.}{Eqs.}
\crefname{figure}{Fig.}{Figs.}
\crefname{table}{Tab.}{Tabs.}
\crefname{section}{Sec.}{Secs.}

\usepackage{amsfonts}
\usepackage{amsmath}
\usepackage{multirow}
\usepackage{amssymb}
\usepackage{amsbsy}
\usepackage{graphicx}
\usepackage{epstopdf}
\usepackage{color}
\usepackage{braket} 
\usepackage{mathdots} 
\usepackage{booktabs}
\usepackage{indentfirst}
\usepackage{diagbox}
\usepackage{lipsum}
\usepackage{mwe}
\usepackage{hyperref}

\hypersetup{colorlinks=true, linkcolor=blue, anchorcolor=blue, citecolor=blue,urlcolor=blue}
\begin{document}
\title{Bosonic Quantum Breakdown Hubbard Model}	

\author{Yu-Min Hu}
\affiliation{Institute for Advanced Study, Tsinghua University, Beijing,  100084, China}
\affiliation{Department of Physics, Princeton University, Princeton, NJ 08544, USA}

\author{Biao Lian}
\affiliation{Department of Physics, Princeton University, Princeton, NJ 08544, USA}
\date{\today}

\begin{abstract}
We propose a bosonic quantum breakdown Hubbard model, which generalizes the Bose-Hubbard model by adding an asymmetric breakdown interaction turning one boson into two between adjacent sites. When the normal hopping is zero, this model has a global exponential U(1) symmetry, and we show that the ground state undergoes a first-order phase transition from a Mott insulator (MI) to a spontaneously symmetry breaking (SSB) breakdown condensate as the breakdown interaction increases. Surprisingly, the SSB breakdown condensate does not have a gapless Goldstone mode, which invalidates the Mermin-Wagner theorem and leads to stable SSB in one dimension. Moreover, we show that the quench dynamics of a boson added to MI exhibits a dynamical transition from dielectric to breakdown phases, which happens at a larger breakdown interaction than the ground state phase transition. Between these two transitions, the MI (dielectric) state is a false vacuum stable against dynamical breakdown. Our results reveal that quantum models with unconventional symmetries such as the exponential symmetry can exhibit unexpected properties.
\end{abstract}
\maketitle

Symmetries are central to the classification of quantum phases of matter. In particular, ground states with spontaneously broken global continuous symmetry are known to host gapless Goldstone modes. A prototypical example is the Bose-Hubbard model \cite{fisher1989boson} for interacting bosons in a lattice, which exhibits a superfluid phase with spontaneously broken global U(1) symmetry and a linear dispersion Goldstone mode. In dimensions $d\le 2$ ($d\le 1$) at finite (zero) temperature, spontaneously broken symmetries are generically restored by quantum fluctuations of the gapless Goldstone modes, according to the Mermin-Wagner theorem \cite{Mermin1966mermin, Hohenberg1967existence}.

In recent years, the study of phases of matter has been extended to various unconventional symmetries. For instance, dipole and multipole symmetries play a vital role in fractonic phases \cite{Nandkishore2019fractons,Pretko2020fracton}, and the spontaneous breaking of dipole and multipole symmetries gives rise to exceptional quantum phases of matter \cite{yuan2020fractonic, chen2021fractonic, Kapustin2022hohenberg, Stahl2022Spontaneous, lake2022dipolar, chen2023manybody, Zechmann2023fractonic, boesl2023deconfinement}. Moreover, many of these unconventional symmetries exert substantial constraints on the non-equilibrium many-body quantum dynamics, such as Hilbert space fragmentation \cite{Khemani2020localization, Sala2020ergodicity, moudgalya2022thermalization, Moudgalya2022hilbert, Moudgalya2022review,kohlert2023exploring} and exotic relaxation hydrodynamics \cite{Feldmeier2020anomalous, zhang2020subdiffusion, ogunnaike2023unifying, Morningstar2023hydrodynamics, gliozzi2023hierarchical, stahl2023fracton, jain2023dipole,Armas2024Ideal}.

Recently, a distinct form of unconventional symmetry, known as exponential symmetry, has garnered attention in various contexts. The exponential U(1) symmetry is generated by an exponential charge $\hat Q=\sum_m q^{-m}\hat n_m$ with a certain number $q\neq 1$ and $\hat n_m$ being the particle number at the $m$-th site. This symmetry has been found to play a significant role in exotic ground states \cite{sala2023exotic, Hu2023spontaneous}, topological phases \cite{Watanabe2023ground,delfino20232d,han2023topological}, and constrained quantum dynamics \cite{lian2023quantum_breakdown,liu20232d, sala2022dynamics,chen2024quantum}.  In particular, in the quantum breakdown model for fermions \cite{lian2023quantum_breakdown}, the exponential U(1) symmetry naturally arises from a spatially asymmetric \emph{breakdown interaction} resembling the electrical breakdown phenomenon, leading to many-body localization.

This motivates us to propose a one-dimensional (1D) bosonic quantum breakdown Hubbard model given in \cref{eq:1D_Hamiltonian}. It generalizes the celebrated Bose-Hubbard model \cite{fisher1989boson} by adding a spatially asymmetric breakdown interaction $J$ which turns one boson in a site into two bosons in the next site (and its conjugate). The model has a global exponential U(1) symmetry when the hopping $\gamma=0$. 

Employing Gutzwiller mean field and density matrix renormalization group (DMRG) methods, we find the ground state of this model at $\gamma=0$ ($\gamma\neq 0$) exhibits a first-order phase transition from a Mott insulator (MI) to breakdown condensate (breakdown insulator) as $J$ increases. At $\gamma=0$, the breakdown condensate spontaneously breaks the global exponential U(1) symmetry, but remarkably gives no gapless Goldstone modes. This invalidates the Mermin-Wagner theorem \cite{Mermin1966mermin, Hohenberg1967existence}, and indicates the spontaneous breaking of exponential U(1) symmetry is robust in any dimension $d>0$. To our knowledge, spontaneous symmetry breaking without Goldstone modes has not been found in regular lattices, although it has been studied in tree graphs \cite{laumann2009}. We further reveal a dynamical phase transition from dielectric to breakdown in the quench dynamics of a boson added to MI as $J$ increases, similar to \cite{lian2023quantum_breakdown}. We find the dynamical phase transition always occurs at a larger $J$ compared to the ground state phase transition; between the two transitions, MI is a false vacuum stable against dynamical breakdown. We also show that our predictions can be observed in a practical experimental proposal. 

\emph{Model.} Inspired by the quantum breakdown model for fermions \cite{lian2023quantum_breakdown}, we propose the \emph{bosonic quantum breakdown Hubbard model} in a 1D lattice of $L$ sites. Each site $m$ has a boson mode with annihilation and creation operators $\hat a_m, \hat a_m^\dagger$. The Hamiltonian is: 
\begin{equation}
    \begin{split}
        H=&-\sum_{m=1}^{L-1}\left[\gamma\hat a_{m+1}^\dagger\hat a_m+ J(\hat a_{m+1}^{\dagger })^2\hat a_m+\text{H.c.}\right] \\
        &+\sum_{m=1}^L\left[-\mu \hat n_m+\frac{U}{2} \hat n_m(\hat n_m-1)\right]\ ,
    \end{split}\label{eq:1D_Hamiltonian}
\end{equation}
where $\hat n_m=\hat a_m^\dagger\hat a_m$ is the boson number on site $m$, and $\text{H.c.}$ represents the Hermitian conjugate. $\gamma\ge 0$ (real) is the nearest neighbor hopping, $\mu$ is the chemical potential, and $U>0$ is the on-site Hubbard interaction. Additionally, there is a spatially asymmetric interaction $J\ge 0$ (real) called the breakdown interaction \cite{lian2023quantum_breakdown}, which induces a progressive proliferation (reduction) of bosons to the right (left) direction. This resembles the Townsend avalanche of numbers of particles (electrons and ions) in the electrical breakdown of dielectric gases. In ultracold atoms, such a breakdown interaction for bosons has been studied in two-site systems, to elucidate the many-body chemical reactions of the formation of diatomic molecules from atomic condensation \cite{zhang2021transition, zhang2023many, Heinzen2000superchemistry, Radzihovsky2004superfluid, Romans2004quantum, Malla2022coherent}. In our model \cref{eq:1D_Hamiltonian}, the Hubbard interaction $U>0$ ensures the total energy is lower bounded. 

\emph{Exponential symmetry.} When $\gamma>0$ and $J>0$, the model has no symmetry other than the discrete translation symmetry $\hat{T}$. However, additional global symmetry exists in the two cases below.

At $J=0$ and $\gamma>0$, the model in \cref{eq:1D_Hamiltonian} reduces to the celebrated Bose-Hubbard model \cite{fisher1989boson}, which has a global U(1) symmetry $\hat a_m\to e^{i\varphi}\hat a_m$ corresponding to the conserved boson number $\hat{N}_\text{tot}=\sum_m \hat n_m$.

At $J>0$ and $\gamma=0$, the model has a symmetry dependent on the boundary condition. For the open boundary condition (OBC) that we assumed in \cref{eq:1D_Hamiltonian}, the model also has a global exponential U(1) symmetry given by
\begin{equation}
\hat a_m\to e^{i\varphi_m}\hat a_m\ , \quad \varphi_m= 2^{L-m}\varphi_L\quad (1\le m\le L), \label{eq:phase_relation}
\end{equation}
with $\varphi_L\in[0,2\pi)$, and the associated conserved U(1) charge is $\hat Q^{(\text{OBC})}=\sum_{m=1}^L2^{L-m}\hat n_m$. If one instead imposes a periodic boundary condition (PBC), \cref{eq:phase_relation} would still be a symmetry if $\varphi_L=2\varphi_1\mod 2\pi$, which requires $\varphi_L=\frac{2\pi p}{2^L-1}$, $p\in \mathbb{Z}$. Thus, the model has a global $\mathbb{Z}_{2^L-1}$ symmetry, and accordingly the conserved charge is $\hat Q^{(\text{PBC})}=\hat Q^{(\text{OBC})}\mod (2^L-1)$. This is similar to other PBC models with exponential symmetry studied recently \cite{Watanabe2023ground, delfino20232d, Hu2023spontaneous}. In the limit $L\rightarrow \infty$, both OBC and PBC effectively have a U(1) symmetry. This U(1) charge $\hat Q$ does not commute with translation, satisfying $\hat T\hat Q=2\hat Q \hat T$, although the Hamiltonian is translationally invariant.

\emph{Ground-state phase diagram.} We now investigate the ground state of our model \cref{eq:1D_Hamiltonian}. Our starting point is $\gamma=J=0$, where the system comprises decoupled sites, and the ground state is a Mott insulator (MI) with an integer number of bosons $\langle \hat n_m\rangle =n\in\mathbb{Z}$ per site when $n-1\le\mu/U\le n$.

At $J=0$, the model reduces to the Bose-Hubbard model \cite{fisher1989boson}, and the mean-field ground state is known to undergo a U(1) spontaneous symmetry breaking (SSB) phase transition from MI to a boson condensate superfluid as $\gamma/U$ increases, characterized by the continuous transition of the order parameter $\phi_m=\braket{\hat a_m}$ from zero to nonzero. The superfluid phase exhibits a gapless Goldstone mode with linear dispersion. By the Mermin-Wagner theorem \cite{Mermin1966mermin, Hohenberg1967existence}, the SSB of superfluid becomes only quasi-long range ordered at zero temperature in 1D, due to quantum fluctuations of the Goldstone modes.

\begin{figure}
\centering
\includegraphics[width=8.5cm]{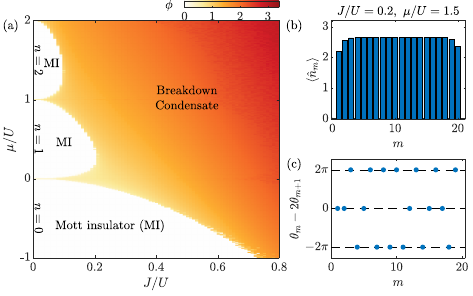}
\caption{(a) The ground-state phase diagram of the model (OBC) in \cref{eq:1D_Hamiltonian} at $\gamma=0$, and we set $U=1$. The colormap shows the order parameter $\phi$.  (b) The ground state density $\braket{\hat n_m}$ and (c) a possible phase angle configuration $\theta_m=\arg[\phi_m]$ in the breakdown condensate ($J/U=0.2$ and $\mu/U=1.5$). The Gutzwiller mean-field ground state is calculated with $N_{\max}=20$ and $L=20$.  }\label{fig:phase_diagram}
\end{figure}

To extend the theory to generic $\gamma$ and $J$, we employ a spatially dependent Gutzwiller mean-field ansatz wavefunction \cite{krauth1992gutzwiller,krutitsky2011excitation} $\ket{\Phi}=\prod_{m=1}^{L}\left(\sum_{n=0}^{N_{\max}}C_{m,n}\frac{(\hat a_m^\dagger)^n}{\sqrt{n!}}\right) \ket{0}$, where $\ket{0}$ is the vacuum state, and we truncate the allowed boson number per site at some large enough $N_{\max}$. We numerically minimize the energy density [for OBC here and PBC in the Supplemental Material (SM) \cite{boson_supp}] $
\mathcal{E}_{\Phi}(\{C_{m,n}\})=\frac{\braket{\Phi|H|\Phi}}{L{\braket{\Phi|\Phi}}}$ with respect to the complex variational parameters $C_{m,n}$, and find the mean field ground state $\ket{\Phi_\text{gs}}$ \cite{boson_supp}. We then define the local complex order parameter $\phi_{m}$ and its mean magnitude $\phi$ as
\begin{equation}\label{eq:order-para}
\phi_{m}=\frac{\braket{\Phi_\text{gs}|\hat a_m|\Phi_\text{gs}}}{\braket{\Phi_\text{gs}|\Phi_\text{gs}}}=|\phi_m|e^{i\theta_m},\quad \phi=\frac{1}{L}\sum_{m=1}^L |\phi_m|.
\end{equation}
We also calculate the boson number $\langle \hat n_m\rangle=\frac{\braket{\Phi_\text{gs}|\hat n_m|\Phi_\text{gs}}}{\braket{\Phi_\text{gs}|\Phi_\text{gs}}}$. At $J=0$, this reproduces the phase diagram of the Bose-Hubbard model \cite{fisher1989boson} with respect to $\mu/U$ and $\gamma/U$.

In the $\gamma=0$ case, which has the exponential U(1) symmetry in \cref{eq:phase_relation}, the ground state phase diagram with respect to $\mu/U$ and $J/U$ is shown in \cref{fig:phase_diagram}. We find the order parameter magnitude $|\phi_m|$ and the filling $\langle \hat n_m\rangle$ are always spatially uniform in the bulk [\cref{fig:phase_diagram}(b)]. At small $J/U$, there are isolated domes of MI phases with $\phi=0$ and integer fillings $\langle \hat n_m\rangle=n\in\mathbb{Z}$ ($n\ge0$). The large $J/U$ regime outside the MI phases develops an order parameter $\phi >0$ and thus exponential U(1) SSB, which we call the \emph{breakdown condensate} phase. In PBC which has $\mathbb{Z}_{2^L-1}$ symmetry, exact diagonalization (ED) verifies that the breakdown condensate has $2^L-1$ degenerate ground states (see SM \cite{boson_supp}). The phase angles $\theta_m=\arg (\phi_m)$ in \cref{eq:order-para} are locked to satisfy $\theta_m=2\theta_{m+1}\mod 2\pi$ as shown in \cref{fig:phase_diagram}(c), which can be rotated into $\theta_m=0$ by the U(1) transformation in \cref{eq:phase_relation}.

\begin{figure}
\centering
\includegraphics[width=8.5cm]{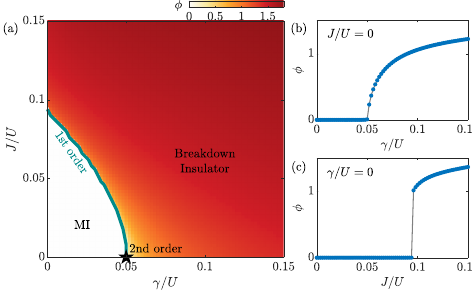}
\caption{(a) The ground state phase diagram at fixed $U=1$ and $\mu/U=1.5$. (b) The order parameter $\phi$ at $J=0$ versus $\gamma/U$ (reduced to the Bose-Hubbard model). (c) The order parameter $\phi$ at $\gamma=0$ versus $J/U$. The calculation is done with $N_{\max}=20$ and $L=20$.}\label{fig:nonzerot}

\end{figure}

Unexpectedly, we find the phase boundary in \cref{fig:phase_diagram}(a) between the MI and the SSB breakdown condensate in the $\gamma=0$ case is always of the first order. This can be seen in \cref{fig:nonzerot}(c), where $\phi$ jumps discontinuously from zero to nonzero at the phase boundary. We further perform DMRG calculations which confirm the first-order phase transition and SSB of the exponential U(1) symmetry (see SM \cite{boson_supp}), despite small quantitative differences due to distinct ansätze. This contrasts with the conventional Bose-Hubbard model ($J=0$), in which $\phi$ undergo second-order phase transitions [\cref{fig:nonzerot}(b)].

In the generic case $\gamma>0$ and $J>0$, the absence of global symmetry other than translation forces the phase boundaries between translationally invariant phases to be of the first order. Indeed, for fixed $\mu$, we find two phases, the MI and breakdown insulator, which are separated by a first-order phase boundary for $\gamma>0$ and $J>0$, as shown in \cref{fig:nonzerot}(a). The bulk of the breakdown insulator has $\phi_m=\phi$ being real and positive. The only second-order phase transition point is the Bose-Hubbard model phase transition on the $\gamma/U$ axis at $J=0$ [\cref{fig:nonzerot}(b)], for which the  breakdown insulator reduces to the superfluid phase.

\emph{Excitations.} To examine the low-energy excitations in the breakdown condensate, we consider the order parameter of the form $\phi_m =\sqrt{\bar{\rho}+\delta\rho_m}e^{i\delta\theta_m}$, with small phase fluctuations $\delta\theta_m$ and density fluctuations $\delta\rho_m$. The Lagrangian for model \cref{eq:1D_Hamiltonian} with OBC reads $\mathcal{L}=\frac{1}{2}\sum_{m=1}^L(i\phi_m^*\partial_t\phi_m-i\phi_m\partial_t\phi_m^*)- \langle H\rangle =\mathcal{L}_0+\delta\mathcal{L}$, where $\mathcal{L}_0$ is the Lagrangian of the mean field ground state with constant particle density $\bar{\rho}=\phi^2>0$. Deep in the breakdown condensate phase, which is well described by a coherent state obeying $\hat a_m|\Phi\rangle=\phi_m|\Phi\rangle$, $\bar{\rho}$ satisfies the saddle-point equation $\mu+2\gamma+3J\sqrt{\bar{\rho}}-U\bar{\rho}=0$, and the Lagrangian fluctuation $\delta\mathcal{L}$ expanded to the second order (up to total derivatives) reads
\begin{equation}
\begin{split}
&\delta\mathcal{L}\approx -\sum_{m=1}^L\Big[\delta\rho_m\partial_t\delta\theta_m+\frac{U}{2}\delta\rho_m^2\Big]\\
&-\sum_{m=1}^{L-1}\Big[J\bar{\rho}^{\frac{3}{2}}(\delta\theta_m-2\delta\theta_{m+1})^2
+\gamma\bar{\rho}(\delta\theta_m-\delta\theta_{m+1})^2\\
&+\frac{J}{4\sqrt{\bar{\rho}}}(\delta\rho_m^2-4\delta\rho_m\delta\rho_{m+1})+\frac{\gamma}{4\bar{\rho}}(\delta\rho_m-\delta\rho_{m+1})^2\Big] \\
&=-\delta \bm{\rho}^T\partial_t\delta\bm{\theta}-\delta\bm{\theta}^T M_\theta \delta\bm{\theta}-\frac{1}{4} \delta \bm{\rho}^T M_\rho \delta \bm{\rho}\ ,
\end{split}
\end{equation}
where we have defined $\delta\bm{\theta}\equiv(\delta\theta_1,\cdots,\delta\theta_L)^T$ and $\delta\bm{\rho}\equiv(\delta\rho_1,\cdots,\delta\rho_L)^T$. The coefficients are rewritten into matrices $M_\theta$ which is non-negative and $M_\rho$ which is positive definite \cite{boson_supp}. Integrating out $\delta\rho_m$ yields an effective Lagrangian $\delta\mathcal{L}_\text{eff}=\partial_t\delta\bm{\theta}^T M_\rho^{-1} \partial_t\delta\bm{\theta}-\delta\bm{\theta}^T M_\theta \delta\bm{\theta}$. Consequently, the Euler-Lagrange equation reads
\begin{equation}\label{eq:excitation}
\partial_t^2\delta\bm{\theta}(t)=-D\delta\bm{\theta}(t)\ ,\qquad D=M_\rho M_\theta\ .
\end{equation}
The excitation energies $\omega$ are square roots of the eigenvalues of the above dynamical matrix $D$.

\begin{figure}
\centering
\includegraphics[width=8.5cm]{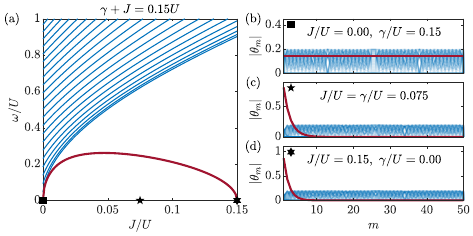}

\caption{(a) The excitation energy spectra of \cref{eq:excitation} in the breakdown condensate/insulator with OBC.  The parameters are fixed at $U=1$, $J+\gamma=0.15U$ and $\mu/U=1.5$.  (b-d) Eigen-wavefunctions of the dynamical matrix $D$ for parameters marked as points in (a). In all panels, bulk states are denoted by blue lines, while the red line denotes the lowest eigenstate which is the only edge mode when $J>0$.}\label{fig:1d_excitations}
\end{figure}

Figure \ref{fig:1d_excitations} shows the excitation spectrum for $L=50$ with OBC at different $J$ and $\gamma$. At $J=0$ and $\gamma>0$, we obtain a linear dispersion gapless Goldstone mode as expected in the conventional superfluid. At $J>0$ and $\gamma\ge0$, we generically find a fully gapped bulk spectrum with $\delta\theta_m$ eigenmodes extended in the bulk [blue thin lines in Figs. \ref{fig:1d_excitations}(a)-\ref{fig:1d_excitations}(d)], and a single in-gap low-lying edge mode with $\delta\theta_m$ exponentially localized at the left edge [red bold lines in Figs. \ref{fig:1d_excitations}(a)-\ref{fig:1d_excitations}(d)]. 

Specifically, when $J>0$ and $\gamma=0$, in the breakdown condensate which spontaneously breaks the exponential U(1) symmetry, the single edge mode of the breakdown condensate reaches zero energy, while the bulk spectrum remains gapped. This bulk gap is further verified by ED with PBC (see SM \cite{boson_supp}). The zero energy edge mode is given by $\delta\theta_m=2^{-m}\delta\theta_0$, which is exactly the exponential U(1) phase rotation. Therefore, there is no gapless Goldstone mode other than the symmetry transformation. Importantly, the absence of gapless excitations invalidates the Mermin-Wagner theorem, and thus we expect the exponential U(1) SSB of the breakdown condensate to survive up to a finite critical temperature $T_c$.

The absence of a gapless Goldstone mode originates from noncommutation of the exponential U(1) charge with translation. By a similarity transformation $\delta\theta_m=2^{-m}\alpha_m$, the exponential U(1) transformation \cref{eq:phase_relation} becomes homogeneously $\alpha_m\rightarrow \alpha_m+\varphi_0$. Taking the continuum limit $\alpha_m(t)\rightarrow\alpha(x,t)$, we find the effective Lagrangian at $\gamma=0$ has the inhomogeneous form $\delta \mathcal{L}_\text{eff}\approx g(x)\left[(\partial_t\alpha)^2-v^2(\partial_x\alpha)^2\right]$, where $g(x)\approx g(0)e^{-2x/\xi}$ with $\xi=\frac{1}{\ln 2}$ and $v>0$. This resembles a massless Klein-Gordon field in a curved spacetime \cite{lv2022curving}, which yields an Euler-Lagrange equation
\begin{equation}
\left(v^{-2}\partial_t^2 -\partial_x^2 +2\xi^{-1}\partial_x\right)\alpha =0\ .
\end{equation}
This corresponds to the celebrated Hatano-Nelson model \cite{Hatano1996localization,Hatano1997Vortex}, a non-Hermitian dynamical matrix that has a gapped real energy spectrum  $\omega=v\sqrt{\xi^{-2}+k^2}$ (with $k$ real) under OBC, with eigenmodes $\alpha(x,t)=\alpha_0 e^{(\xi^{-1}+ik)x-i\omega t}$ exhibiting the non-Hermitian skin effect \cite{yao2018edge}. When transformed back to the $\delta\theta_m$ basis, such eigenmodes are bulk plane waves $\delta\theta(x,t)=\delta\theta_0 e^{ikx-i\omega t}$. The zero energy mode $\alpha(x,t)=\alpha_0$ corresponds to the edge mode in the $\delta\theta_m$ basis.

\emph{Dynamical breakdown.} The fermionic quantum breakdown model in Ref. \cite{lian2023quantum_breakdown} exhibits a transition resembling the electrical breakdown in the quench of single-fermion states. For the boson model here, the existence of MIs, which resemble dielectrics, allows us to explore the breakdown transition of MI versus $J$ under minimal local perturbations, generalizing Ref. \cite{lian2023quantum_breakdown}. Specifically, we set $\gamma=0$ [exponential U(1) symmetric], and define the fixed-point $n$-boson-per-site MI state $|\text{MI},n\rangle=\prod_{m=1}^L\frac{(\hat a_m^\dagger)^n}{\sqrt{n!}}|0\rangle$. We then examine the quench dynamics of the initial state $|\Psi_0^{(n)}\rangle=\frac{\hat a^\dagger_1}{\sqrt{n+1}}|\text{MI},n\rangle$, which perturbs the MI by adding one boson to site $1$. We take OBC and perform exact diagonalization (ED) in the corresponding charge $\hat Q^{\text{OBC}}$ sector, which is finite dimensional for finite lattice size $L$. This gives the boson number $\langle \hat n_m(t)\rangle=\langle\Psi_0^{(n)}|e^{iHt}\hat n_m e^{-iHt}|\Psi_0^{(n)}\rangle$ at each site $m$ with respect to time $t$, and its long-time average $\overline{n}_m=\lim_{T\rightarrow\infty}\frac{1}{T}\int_0^T dt\langle \hat n_m(t)\rangle=\sum_{j}|\braket{e_j|\Psi_0^{(n)}}|^2\braket{e_j|\hat n_m|e_j}$ in terms of the energy eigenstates $|e_j\rangle$ (which is accurate given no accidental degeneracy).

\begin{figure}
\centering
\includegraphics[width=8.5cm]{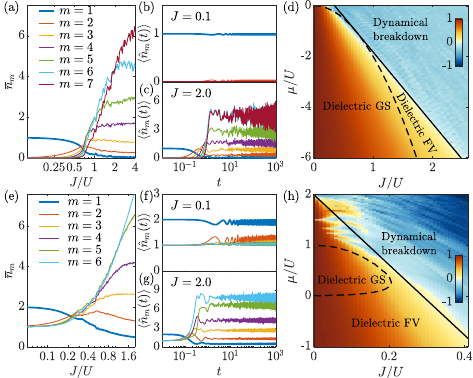}
\caption{The dynamical breakdown transition of initial states $\ket{\Psi_0^{(0)}}$ for $L=7$ in (a)-(d) and $\ket{\Psi_0^{(1)}}$ for $L=6$ in (e)-(h). We fix $U=1$, $\gamma=0$. (a), (e) shows the time-averaged boson numbers $\overline{n}_m$ on site $m$ with respect to $J$, (b), (c), (f), and (g) show the boson number time evolutions $\braket{\hat n_m(t)}$ for $J=0.1$ and $J=2.0$, where $\mu=-1$ in (a)-(c) and $\mu=0.5$ in (e)-(g). In (d) and (h), the colormap shows $\Delta\overline{n}_{12}=\overline{n}_1-\overline{n}_2$ for $\ket{\Psi_0^{(0)}}$ and $\ket{\Psi_0^{(1)}}$, respectively; the solid line plots Eq.\eqref{eq:negative_boundary} with $n=0$ and $n=1$, and the dashed line indicates the ground state phase boundary of the $n=0$ MI and $n=1$ MI (from \cref{fig:phase_diagram}(a)), respectively. GS (FV) implies the dielectric phase is the ground state (false vacuum).  }
\label{fig:dynamics}

\end{figure}

In the breakdown phase, the boson added to site $1$ will evolve into many bosons on sites $m>1$, leading to $\overline{n}_1\to  n$ or $\overline{n}_1<n$. In the dielectric (i.e., no breakdown) phase, the added boson will be trapped, and thus $\overline{n}_1-n$ remains finitely positive. Figure \ref{fig:dynamics} shows the ED results for $\ket{\Psi_0^{(0)}}$ with $L=7$ and $\ket{\Psi_0^{(1)}}$ with $L=6$. The dynamical breakdown transition (subject to finite size effect) can be clearly identified in Figs. \ref{fig:dynamics}(a) and \ref{fig:dynamics}(e), where $\overline{n}_{m>1}$ exceeds $\overline{n}_1$. Figures \ref{fig:dynamics}(b), \ref{fig:dynamics}(c), \ref{fig:dynamics}(f), and \ref{fig:dynamics}(g) shows the distinct $\langle \hat n_m(t)\rangle$ time evolutions in the dielectric and breakdown phases. In Figs. \ref{fig:dynamics}(d) and \ref{fig:dynamics}(h), we calculate $\Delta\overline{n}_{12}=\overline{n}_1-\overline{n}_2$, and identify the $\Delta\overline{n}_{12}>0$ ($\Delta\overline{n}_{12}<0$) regime as the dielectric (breakdown) phase.

The breakdown transition can be estimated as follows. By hopping the added boson on site $1$ into two bosons on site $2$, the system gains a hopping energy $2|\langle\text{MI},n|\frac{(\hat a_2)^2}{\sqrt{(n+1)(n+2)}} H |\Psi_0^{(n)}\rangle|=2(n+1)\sqrt{n+2}J$, while costs an on-site energy $(n+1)U-\mu$. The breakdown transition happens when the energy gain and cost are equal:
\begin{equation}\label{eq:negative_boundary}
2(n+1)\sqrt{n+2}J=(n+1)U-\mu\ .
\end{equation}
This gives the solid lines in Figs. \ref{fig:dynamics}(d) and \ref{fig:dynamics}(h), which agree well with the transition identified by $\Delta\overline{n}_{12}=0$.

We observe that for each MI state $|\text{MI},n\rangle$, the dynamical breakdown transition [\cref{eq:negative_boundary}, solid lines in Figs. \ref{fig:dynamics}(d) and \ref{fig:dynamics}(h)] is always at larger $J$ than the ground state phase boundary of the MI in \cref{fig:phase_diagram}(a) (dashed lines in Figs. \ref{fig:dynamics}(d) and \ref{fig:dynamics}(h)]. This indicates dynamical breakdown cannot happen when the MI is the ground state, since bosons are localized. Interestingly, between the dashed and solid lines in Figs. \ref{fig:dynamics}(d) and \ref{fig:dynamics}(h) where the ground state becomes the breakdown condensate after the first order phase transition, our results imply that MI is still a \emph{false vacuum} stable against dynamical breakdown, as protected by the bulk gap of breakdown condensate.

Furthermore, we find that the level spacing statistics of these charge sectors show a crossover from Poisson to Wigner-Dyson as $J/U$ increases \cite{boson_supp}. Thus, the model is quantum chaotic and quickly thermalizes in the breakdown phase. This agrees with Figs. \ref{fig:dynamics}(c) and \ref{fig:dynamics}(g) in the breakdown phase, where the boson on site $1$ decays and reaches equilibrium in a time scale $\sim J^{-1}$.

 \emph{Possible experiments.} The seemingly unusual breakdown interaction in Eq.\eqref{eq:1D_Hamiltonian}  can be implemented in cold-atom experiments. Since the breakdown interaction does not preserve particle number, we introduce a driven ancilla boson mode per site with creation operator $\hat b_m^\dagger$ as a particle source. A practical Hamiltonian is  $H=-\sum_{m=1}^{L-1}[J_da_{m}(\hat a_{m+1}^{\dagger })^2\hat b_{m+1}+\text{h.c.}]+\sum_{m=1}^L[-\mu \hat a_m^
\dagger a_m+\frac{U}{2} \hat a_m^
\dagger \hat a_m^
\dagger a_m a_m]+\sum_{m=2}^{L}[-\mu \hat b_m^
\dagger b_m+\frac{U}{2} \hat b_m^
\dagger \hat b_m^
\dagger b_m b_m+F(b_{m}+b_m^\dagger)]$. Here, $F$ is the driving (made time-independent after a rotating wave transformation) that pumps bosons into the system, and $J_d$ is a local-dipole-conserving interaction which can be experimentally implemented with tilted potentials \cite{Yang2020observation,Scherg2021Observing,su2023observation}.  With $F\ne0$, this Hamiltonian has the global exponential U(1) symmetry of Eq. \eqref{eq:phase_relation}. A nonzero $F$ drives the ancilla bosons into a condensate, and the Gutzwiller mean-field calculation presented in the SM \cite{boson_supp} demonstrates a bulk translational invariant mean field $\braket{\hat b_m}=\phi_b$. This effectively leads to the bosonic quantum breakdown Hubbard model in Eq.\eqref{eq:1D_Hamiltonian} with $J= J_d\phi_b$ and $\gamma=0$.  In the SM \cite{boson_supp}, we show that the mean-field phase diagram for this practical model is similar to Fig.\ref{fig:phase_diagram}(a). 

\emph{Discussions.} We showed that the 1D bosonic quantum breakdown Hubbard model at $\gamma=0$, which has a global exponential U(1) symmetry in \cref{eq:phase_relation}, undergoes a ground state phase transition from MI to the SSB breakdown condensate, but gives no gapless Goldstone mode. This raises questions about the non-relativistic Goldstone theorem \cite{watanabe2012} and Mermin-Wagner theorem for generic continuous symmetries. The absence of Goldstone modes may also be connected with the physics of non-Hermitian models \cite{fukui1998,oka2010,lv2022curving} and models on trees \cite{laumann2009}. The algebra $\hat T\hat Q=2\hat Q\hat T$ between the exponential U(1) charge $\hat Q$ and translation $\hat T$ occurs in the $q$-deformed harmonic oscillator and quantum group \cite{sun1989,Biedenharn_1989,Macfarlane_1989}, which may yield a deeper understanding of the above. The dynamical breakdown transition happens at a larger $J$ than the ground state phase transition, between which the MI (dielectric) phase is a false vacuum. This provides a prototypical example for studying the lifetime of false vacuum and quantum scars \cite{yin2023}. It will also be interesting to explore the spin analogs of our model (by the Holstein–Primakoff transformation, etc.), and their realization in experimental platforms such as ultracold atoms  \cite{zhang2021transition,zhang2023many}. Lastly, the exponentially many degenerate gapped ground states ($2^L-1$ in PBC) in the breakdown condensate, combined with the analogy between the exponential charge $\hat{Q}=\sum_{m=1}^L2^{L-m}\hat n_m$ and binary numbers, may find intriguing applications in quantum memory and quantum algorithm implementations.

\begin{acknowledgments}
\emph{Acknowledgements.} We thank Bo-Ting Chen, Hao Chen,  Yichen Hu, Zhou-Quan Wan, Shunyu Yao, Wucheng Zhang, Ping Gao, and Zheng-Cheng Gu for helpful discussions. Y.-M. Hu acknowledges the support from NSFC under Grant No. 12125405 and the Tsinghua Visiting Doctoral Students Foundation. This work is supported by the Alfred P. Sloan Foundation, the National Science Foundation through Princeton University’s Materials Research Science and Engineering Center DMR-2011750, and the National Science Foundation under award DMR-2141966. Additional support is provided by the Gordon and Betty Moore Foundation through Grant GBMF8685 towards the Princeton theory program.
\end{acknowledgments}

\bibliography{boson_ref}

\end{document}


\title{The supplemental material for “Bosonic Quantum Breakdown Hubbard Model"}	

\author{Yu-Min Hu}
\affiliation{Institute for Advanced Study, Tsinghua University, Beijing,  100084, China}
\affiliation{Department of Physics, Princeton University, Princeton, NJ 08544, USA}

\author{Biao Lian}
\affiliation{Department of Physics, Princeton University, Princeton, NJ 08544, USA}

\maketitle
\section{DMRG calculations of the ground states}
In this section, we numerically study the ground-state properties of the bosonic quantum breakdown Hubbard model using the density matrix renormalization group (DMRG) algorithm, implemented with the TeNPy package \cite{hauschild2024tensornetworkpythontenpy}. The DMRG results qualitatively confirm the ground-state properties obtained from the variational mean-field approach in the main text.

For simplicity, in the DMRG simulation, we fix the model parameters as $U=1$, $\gamma=0$, $\mu=\pm0.5$, and $L=20$. We then increase the value of the breakdown interaction $J$ to explore the possibility of a phase transition. For a relatively small $J$, the dimension of the truncated local Hilbert space is set to $D=11$, which means that the maximum number of bosons per site is 10. Additionally, the bond dimension of the matrix product states is set to $\chi=64$. In the thermodynamic limit, the spontaneous symmetry-breaking phase exhibits a massive degeneracy of ground states. This degeneracy is lifted by small energy gaps in a finite system, making the DMRG algorithm prone to convergence toward local minima rather than the true ground state. To mitigate this finite-size effect, we initialize the DMRG algorithm with several random initial states, compute the ground state for each, and select the one with the lowest energy as the ground-state energy.

\begin{figure}
    \centering
    \includegraphics[width=0.75\linewidth]{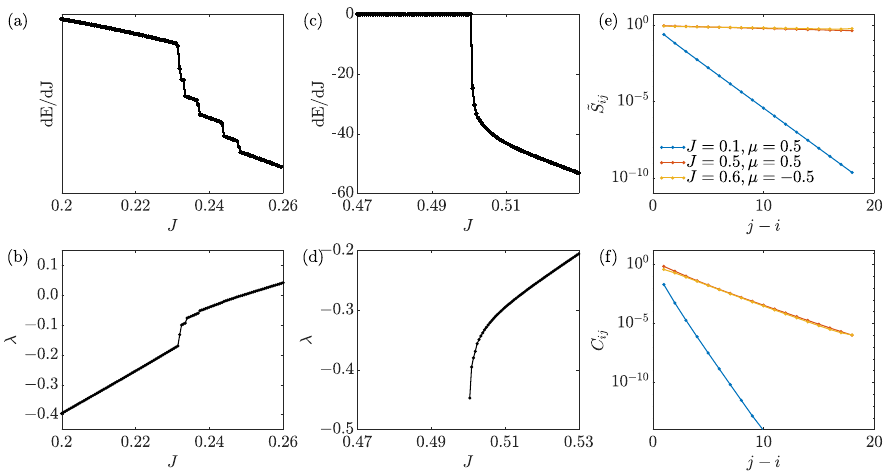}
    \caption{Ground state properties obtained by the DMRG algorithm. (a,c) $\frac{\mathrm{d}E}{\mathrm{d}J}$ with the change of $J$. (b,d) The exponent of $S_{ij}$ in Eq.\eqref{seq:S_exponent}. (e)  $\tilde S_{ij}$ in Eq.\eqref{seq:renormalized_S} with $i=1$. (f)  $C_{ij}$ in Eq.\eqref{seq:nn_correl} with $i=1$. We set the chemical potential $\mu=0.5$ in (a,b) and $\mu=-0.5$ in (c,d). Note that $\lambda$ in (d) is not well-defined in the Mott insulating region where the ground state is exactly the vacuum state $\ket{0}$. Other parameters are $\gamma=0$, $U=1$, $L=20$.}
    \label{fig:dmrgv2}
\end{figure}

We begin by calculating the ground-state energy \(E(J)\) as a function of the interaction strength \(J\), presenting numerical results for both positive and negative chemical potentials. As shown in Fig.~\ref{fig:dmrgv2}(a,c), the first derivative of the energy, \(\frac{\mathrm{d}E}{\mathrm{d}J}\), exhibits a sharp jump around \(J \approx 0.23\) for \(\mu = 0.5\) and \(J \approx 0.50\) for \(\mu = -0.5\), providing clear evidence of a first-order phase transition. We note that the critical point for \(\mu = 0.5\) differs slightly from the mean-field result (\(J_c \approx 0.18\) for \(\mu = 0.5\) and \(U = 1\)), as shown in Fig.~1(a) of the main text. This discrepancy arises from the different wavefunction ansätze employed in the two numerical approaches. Despite this quantitative difference, both methods consistently indicate the occurrence of a first-order phase transition driven by increasing breakdown interactions.

To further characterize the nature of phase transition, we calculate the correlation functions of the ground states. In the conventional Bose-Hubbard model, we often employ the two-point correlation function $\braket{\hat a_i \hat a_j^\dagger}$ to detect the off-diagonal long-range order in the superfluid phase. An analog in the bosonic  quantum breakdown Hubbard model  might be the two-point correlation function $\braket{\hat a_i(\hat a_j^\dagger)^{2^{j-i}}}$. However, such a two-point correlation function is not feasible for practical use, since $(\hat a_j^\dagger)^{2^{j-i}}$ with a large $j-i$ becomes ill-defined in a truncated local Hilbert space. Instead, we propose a string operator $\hat a_i\prod_{m=i+1}^{j-1}\hat a_m^\dagger (\hat a_j^\dagger)^2$ that commutes with the exponential charge. The expectation values of the string operator are given by:
\begin{equation}
    S_{ij}=\braket{\psi|\hat a_i\prod_{m=i+1}^{j-1}\hat a_m^\dagger (\hat a_j^\dagger)^2|\psi}
\end{equation}
where we always assume $j>i$. In the above expression, $\ket{\psi}$ is the ground-state wavefunction obtained by the DMRG algorithm. Numerically, we find that $S_{ij}$ scales exponentially with respect to $j-i$:
\begin{equation}\label{seq:S_exponent}
    S_{ij}\propto e^{\lambda(j-i)}.
\end{equation}

As shown in Fig.~\ref{fig:dmrgv2}(b) for \(\mu = 0.5\), the exponent \(\lambda\) exhibits a discontinuous jump around \(J \approx 0.23\), providing further evidence for a first-order phase transition. Additionally, we observe that \(\lambda\) remains negative throughout the Mott insulating phase (\(J < 0.23\)), but becomes positive in the breakdown condensate phase (\(J > 0.23\)). This indicates that \(S_{ij}\) may grow exponentially when the breakdown interaction \(J\) becomes sufficiently large. On the other hand, for \(\mu = -0.5\), the ground state in the Mott insulating phase is exactly the vacuum state \(\ket{0}\), rendering the exponent \(\lambda\) ill-defined for \(J < J_c \approx 0.50\). Consequently, Fig.~\ref{fig:dmrgv2}(d) only presents numerical values of \(\lambda\) for \(J > J_c\). Since the vacuum state \(\ket{0}\) remains an exact eigenstate even when \(J > J_c\), the transitions shown in Fig.~\ref{fig:dmrgv2}(c,d) clearly demonstrate an energy level crossing, which is a hallmark of a first-order phase transition.

Deep in the breakdown condensate phase, at the mean-field level, we expect that $\braket{\hat b_m} = \sqrt{n_m} e^{i 2^{L-m} \varphi}$, where $n_m = \braket{\psi |\hat  a_m^\dagger \hat a_m | \psi}$ represents the particle number at the $m$th site, and $\varphi$ is the phase angle associated with the symmetry-breaking mean-field ground state. In this context, the expectation value of the string operator is expected to be $\braket{\psi |\hat a_i \prod_{m=i+1}^{j-1} \hat a_m^\dagger (\hat a_j^\dagger)^2 | \psi} \approx n_j \prod_{m=i}^{j-1} \sqrt{n_m}$. In contrast, due to the lack of phase coherence in the Mott insulator phase, the expectation value of the string operator should vanish as $ j - i $  increases. 

Based on the mean-field expectations, we anticipate that the renormalized quantity
\begin{equation}\label{seq:renormalized_S}
    \tilde{S}_{ij} = \frac{S_{ij}}{n_j \prod_{m=i}^{j-1} \sqrt{n_m}}
\end{equation}
will display different behaviors in the two phases. As illustrated in Fig. \ref{fig:dmrgv2}(e), $\tilde{S}_{ij}$ demonstrates an exponential decay within the Mott insulator phase, suggesting the absence of long-range phase coherence. In contrast, in the breakdown condensate phase, $\tilde{S}_{ij}$ remains nearly constant, which signifies the presence of long-range phase coherence in this phase and highlights the spontaneous breaking of the exponential U(1) symmetry\footnote{An extremely slow decrease of $\tilde{S}_{ij}$ is observed numerically, which arises from quantum correlations that extend beyond the mean-field approximation.}.

For comparison, we also present the density-density correlation function:
\begin{eqnarray}\label{seq:nn_correl}
    C_{ij} = \braket{\psi | \hat{n}_i \hat{n}_j | \psi} - \braket{\psi | \hat{n}_i | \psi} \braket{\psi | \hat{n}_j | \psi}.
\end{eqnarray}
As illustrated in Fig. \ref{fig:dmrgv2}(f), $C_{ij}$ decreases exponentially  for $J = 0.1$ and $\mu=0.5$, aligning with the gapped Mott insulator phase. More interestingly, $C_{ij}$ also shows an exponential decay for $J = 0.5,\mu=0.5$ and $J=0.6,\mu=-0.5 $. This finding implies the existence of a gapped excitation spectrum beyond the symmetry-breaking ground state within the breakdown condensate phase.

In summary, the DMRG calculations corroborate the ground-state properties predicted by the variational Gutzwiller mean-field approach discussed in the main text. The mean-field method was employed in the main text because it provides a clear demonstration of the phase relation ($\theta_m = 2\theta_{m+1} \mod 2\pi$) of the breakdown condensate. This explicit representation facilitates a straightforward understanding of the spontaneous breaking of the exponential U(1) symmetry. 

We additionally note that the DMRG phase boundary ($J_c\approx0.23$ for $\mu=0.5$ and $U=1$, see Fig.\ref{fig:dmrgv2}(a)) appears slightly larger than the transition point for dynamical breakdown from Eq. (7) of the main text ($J_c\approx0.22$ where $n=1$, $U=1$, and $\mu=0.5$), unlike the mean field phase boundary which is smaller than the dynamical breakdown transition point. However, it remains reasonable to expect the existence of a false vacuum between the dielectric MI ground state and the dynamical breakdown phase, since the MI has all bosons localized. The expectation stems from the fact that the approximate estimation that leads to Eq. (7) in the main text is based on the quench dynamics initiated by exciting a boson at the first site in a MI state only accurately defined at $J=0$ (which is also the mean field ground state). This may differ from quench dynamics initiated by a local excitation in the MI ground state at finite interaction $J\rightarrow J_c$, which we expect only exhibits dynamical breakdown when the MI is no longer the ground state ($J>J_c$).

\section{Coefficient matrices in the effective Lagrangian}
In the main text, we have derived an effective Lagrangian to describe the low-energy phase fluctuations of the breakdown condensate. This Lagrangian is given by
\begin{eqnarray}
\delta\mathcal{L}_\text{eff}=\partial_t\delta\bm{\theta}^T M_\rho^{-1} \partial_t\delta\bm{\theta}-\delta\bm{\theta}^T M_\theta \delta\bm{\theta}.
\end{eqnarray}

In this section, we provide explicit formulations for the coefficient matrices $M_\rho$ and $M_\theta$ of dimension $L\times L$, with $L$ being the lattice length. These expressions are extracted directly from Eq. (5) in the main text. With $\mathbb{I}$ being the identity matrix of dimension $L$, the expressions are given by
\begin{widetext}
\begin{eqnarray}
\begin{split}
    M_\rho&=2U\mathbb{I}+\frac{J}{ \sqrt{\bar{\rho}}}\left(\begin{array}{ccccc}
1 & -2 & & & \\
-2 & 1 & -2 & & \\
& \ddots & \ddots & \ddots & \\
& & -2 & 1 & -2 \\
& & & -2 & 0
\end{array}\right)+\frac{\gamma}{ \bar{\rho}}\left(\begin{array}{ccccc}
1 & -1 & & & \\
-1 & 2 & -1 & & \\
& \ddots & \ddots & \ddots & \\
& & -1 & 2 & -1 \\
& & & -1 & 1
\end{array}\right), \\
M_\theta&=J{ \bar{\rho}}^{\frac{3}{2}}\left(\begin{array}{ccccc}
1 & -2 & & & \\
-2 & 5 & -2 & & \\
& \ddots & \ddots & \ddots & \\
& & -2 & 5 & -2 \\
& & & -2 & 4
\end{array}\right)+{\gamma}{\bar{\rho}}\left(\begin{array}{ccccc}
1 & -1 & & & \\
-1 & 2 & -1 & & \\
& \ddots & \ddots & \ddots & \\
& & -1 & 2 & -1 \\
& & & -1 & 1
\end{array}\right).
\end{split}
\end{eqnarray}
\end{widetext}

\section{The level spacing statistics in some charge sectors}
 
In this section, we employ exact diagonalization techniques to study the level spacing statistics of the $\gamma=0$ bosonic quantum breakdown Hubbard model. Specifically, we focus on the charge sectors generated by $\ket{\Psi_0^{(0)}}$ and $\ket{\Psi_0^{(1)}}$, respectively. These two reference states are used to study the dynamical breakdown transition in the main text.

\begin{figure*}
     \centering     
     \includegraphics[width=\textwidth]{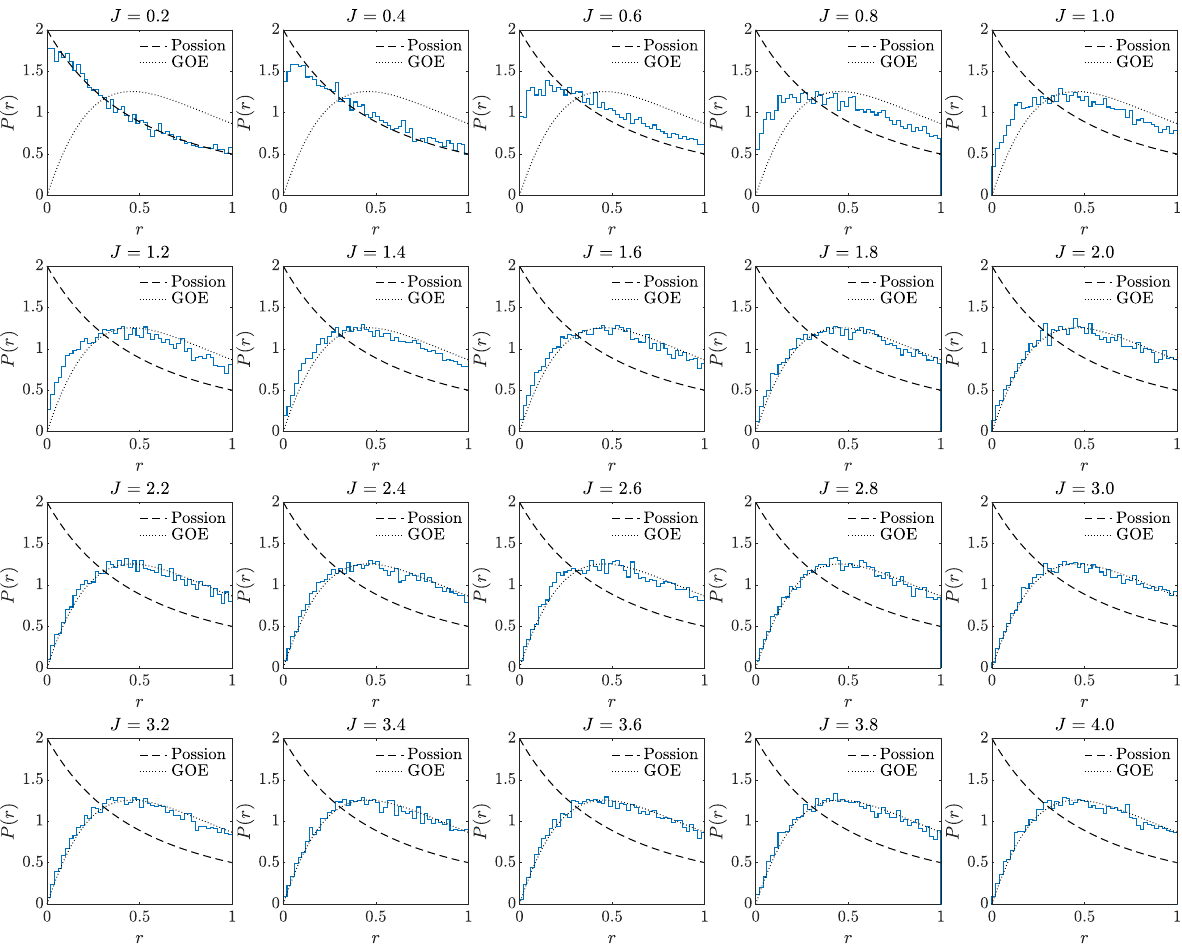}     
     \caption{The distribution $P(r)$ of $r_n$ for varying values of $J/U$ in the charge sector of $\ket{\Psi_0^{(0)}}$, with fixed parameters set as $U=1$, $\gamma=0$, $\mu=-1$, and $L=8$.}
     \label{sfig:distribution}
 \end{figure*}

\begin{figure}
     \centering
     \includegraphics{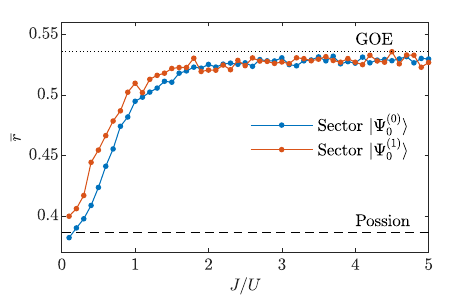}
     \caption{The mean value $\overline{r}$ of the charge sectors generated by $\ket{\Psi_0^{(0)}}$ and $\ket{\Psi_0^{(1)}}$, respectively. For the sector $\ket{\Psi_0^{(0)}}$, we set $U=1$, $\gamma=0$, $\mu=-1$, and $L=8$; For the sector $\ket{\Psi_0^{(1)}}$, we set $U=1$, $\gamma=0$, $\mu=0.5$, and $L=6$.}
     \label{sfig:r_change}
 \end{figure}

To study the statistics of level spacing, we calculate the ratio of consecutive level spacings and then compare the resulting distribution with the established outcomes of the Poisson and Wigner-Dyson distributions. Assuming that $E_n$ is the $n$-th energy level of the ordered energy spectrum in a charge sector,  we compute the quantity
\begin{eqnarray}
    r_n=\frac{\min(s_{n-1},s_n)}{\max(s_{n-1},s_n)},
\end{eqnarray}
where $s_n=E_{n+1}-E_n$. Then the distributions of $r_n$ for different choices of $J/U$ are compared with the results of the random matrix theory \cite{Oganesyan2007localization,Atas2013distribution}. As shown in Fig. \ref{sfig:distribution}, as the parameter $J/U$ increases, the distribution of $r_n$ transitions from a Poisson distribution to a Wigner-Dyson distribution of the Gaussian orthogonal ensemble (GOE). This crossover is also manifested in the mean value $\overline{r}=\braket{r_n}$, which undergoes an increase from the Poisson value 0.386 to the GOE value 0.536 [Fig. \ref{sfig:r_change}]. These results indicate that the model is quantum chaotic when $J$ is sufficiently large.

\section{Mean-field calculation of the phase diagram}\label{apsec:meanfield}
In the main text, we employ the spatially dependent Gutzwiller wavefunction to variationally find the mean-field ground state. As introduced in the main text, the wavefunction ansatz is given by 
\begin{eqnarray}
\ket{\Phi}=\prod_{m=1}^{L}\left(\sum_{n=0}^{N_{\max}}C_{m,n}\frac{(\hat a_m^\dagger)^n}{\sqrt{n!}}\right) \ket{0}\ .\label{seq:GW_general}
\end{eqnarray}
Under this assumption, the local complex coherent field $\phi_{a,m}=\braket{\hat a_m}$  can be expressed as
\begin{eqnarray}
    \phi_{a,m}=\frac{\braket{\Phi|\hat a_m|\Phi}}{\braket{\Phi|\Phi}}=\frac{\sum_{n=0}^{N_{\max}-1} C_{m,n}^* C_{m,n+1} \sqrt{n+1}}{\sum_{n=0}^{N_{\max}}\left|C_{m,n}\right|^2}.
\end{eqnarray}
Similarly, the local pairing function $\phi_{aa,m}=\braket{\hat a_m\hat a_m}$ is given by
\begin{eqnarray}
\begin{split}
    \phi_{aa,m}=\frac{\braket{\Phi|\hat a_m\hat a_m|\Phi}}{\braket{\Phi|\Phi}}=\frac{\sum_{n=0}^{N_{\max}-2} C_{m,n}^* C_{m,n+2} \sqrt{(n+1)(n+2)}}{\sum_{n=0}^{N_{\max}}\left|C_{m,n}\right|^2}.
\end{split}
\end{eqnarray}
The expression is similar for the density operator:
\begin{eqnarray}
    \braket{\hat n_m}=\frac{\braket{\Phi|\hat a_m^\dagger\hat a_m|\Phi}}{\braket{\Phi|\Phi}}=\frac{\sum_{n=0}^{N_{\max}}n\left|C_{m,n}\right|^2}{\sum_{n=0}^{N_{\max}}\left|C_{m,n}\right|^2}.
\end{eqnarray}
As a result, the energy density of this wavefunction is 
\begin{eqnarray}
\begin{split}
    \mathcal{E}_\Phi(\{&C_{m,n}\})=\frac{\braket{\Phi|H|\Phi}}{L\braket{\Phi|\Phi}}\\
    =&\frac{1}{L}\sum_{m=1}^L\frac{\sum_{n=0}^{N_{\max}}[-\mu n+Un(n-1)/2]\left|C_{m,n}\right|^2}{\sum_{n=0}^{N_{\max}}\left|C_{m,n}\right|^2}\\
    &-\frac{\gamma}{L}\sum_{m=1}^{L-1}(\phi_{a,m}\phi_{a,m+1}^*+\phi_{a,m}^*\phi_{a,m+1})\\
    &-\frac{J}{L}\sum_{m=1}^{L-1}(\phi_{a,m}\phi_{aa,m+1}^{*}+\phi_{a,m}^*\phi_{aa,m+1}).
    \end{split}
\end{eqnarray}

Numerically minimizing this energy function with respect to the complex variational parameters $C_{m,n}$ provides the ground-state phase diagram of the $\gamma=0$ bosonic quantum breakdown Hubbard model, which is shown in Fig.1(a) of the main text. 

As for Fig. 2(a) in the main text, the optimization of the energy function involves three different assumptions of the variational parameter space: real positive, real, and complex variational parameters. The final results are then selected from these three outcomes by comparing the resulting ground-state energies. This comparison approach effectively mitigates the potential convergence to local minima above the global minimum of the energy function. In particular, when $J>0$ and $\gamma>0$, the results restricted in the space of real positive variational parameters produce the minimum energy of the breakdown insulator phase. This observation indicates that in the breakdown insulator phase, $\phi_m>0$ is real positive.

 \section{Bosonic quantum breakdown Hubbard model under periodic boundary conditions}
In this section, we present the mean-field analysis of the bosonic quantum breakdown Hubbard model under periodic boundary conditions (PBC). With $\hat a_m=\hat a_{m+L}$, the PBC Hamiltonian for $\gamma=0$ is
 \begin{equation}
        H=-\sum_{m=1}^{L}\left[ J(\hat a_{m+1}^{\dagger })^2\hat a_m+\text{h.c.}\right]+\sum_{m=1}^L\left[-\mu \hat n_m+\frac{U}{2} \hat n_m(\hat n_m-1)\right]
        \label{seq:PBC_Hamiltonian}
\end{equation}

\begin{figure}[t]
    \centering
    \includegraphics[width=\linewidth]{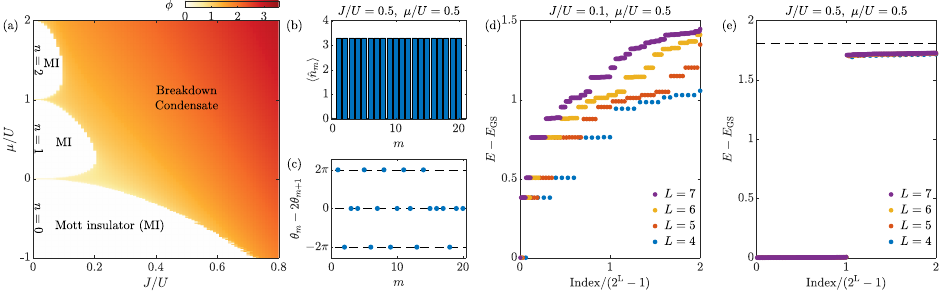}
    \caption{(a) The mean-field ground-state phase diagram of the bosonic quantum breakdown Hubbard model under periodic boundary conditions ($\gamma=0$). The Gutzwiller mean-field ground state is calculated with $N_{\max}=20$, $L=20$, and $U=1$. The colormap shows the order parameter $\phi=\frac{1}{L}\sum_{m=1}^L|\braket{\hat a_m}|$.  (b) The mean-field ground-state density $\braket{\hat n_m}$ and (c) a possible phase angle configuration $\theta_m=\arg[\phi_m]$ in the breakdown condensate ($J/U=0.5$ and $\mu/U=0.5$). (d) and (e) illustrates the low-energy part of the exact-diagonalization spectrum under periodic boundary conditions, where the index of each eigenvalue is obtained in an ascending order. These data points are obtained from exactly diagonalizing the periodic Hamiltonian at each exponential charge section with the system size $L=4,5,6,7$ and an onsite boson number cutoff $N_{\max}=10$. We set $U=1,\ \mu=0.5$ in both plots, while choosing $J=0.1$ in (d) and $J=0.5$ in (e).  The energy shift $E_{\text{GS}}$ is the ground-state energy. The numerical results in (e) are overlapping for different $L$. The dashed line in (e) represents the bulk gap obtained from mean-field expansion (see Fig. 3 of the main text).}
    \label{sfig:pbc}
\end{figure}

As discussed in the main text, the PBC Hamiltonian has a $\mathbb{Z}_{2^L-1}$ symmetry, which can be approximated as a U(1) symmetry in the thermodynamic limit. Following the Gutzwiller mean-field method introduced in Sec. \ref{apsec:meanfield}, we obtain the mean-field phase diagram [Fig. \ref{sfig:pbc}(a)] for the order parameter $\phi=\frac{1}{L}\sum_{m=1}^L|\braket{\hat a_m}|$, which is similar to the mean-field phase diagram under open boundary conditions [Fig. 1(a) of the main text].  Meanwhile, we find that in the breakdown condensate phase, the mean-field ground-state density $\braket{\hat n_m}=\braket{\hat a_m^\dagger \hat a_m}$ and the amplitude of $\phi_m=\braket{\hat a_m}$  are uniformly distributed [Fig. \ref{sfig:pbc}(b)]. More interestingly, the phase angles $\theta_m=\arg\phi_m$ satisfy a phase relation $\theta_m-2\theta_{m+1}=0\mod 2\pi$ [Fig. \ref{sfig:pbc}(c)].   These results indicate the spontaneous breaking of the discrete $\mathbb{Z}_{2^L-1}$ symmetry, which asymptotically becomes the spontaneous breaking of the exponential U(1) symmetry in the thermodynamic limit. 

We further employ exact diagonalization (ED) to investigate the low-energy spectrum of this PBC Hamiltonian. In a finite system of length $L$, the generator of $\mathbb{Z}_{2^L-1}$ symmetry is $\hat Q^{(\text{PBC})}=\sum_{m=1}^L2^{L-m}\hat n_m \mod (2^L-1)$, whose eigenvalues label $2^{L}-1$ distinct symmetry sectors.  With a boson number cutoff $N_{\max}=10$ at each site, we employ ED to obtain the low-energy spectrum in each symmetry sector and sort them in ascending order. The low-energy spectrum of the PBC Hamiltonian is shown in Fig. \ref{sfig:pbc}(d) for the parameter in the Mott insulating phase and in Fig. \ref{sfig:pbc}(d) for the parameter in the breakdown condensate phase. While the spectrum in the Mott insulating phase has a unique symmetric ground state, the ground states in the breakdown condensate phase have a degeneracy of $2^L-1$, indicating the spontaneous breaking of $\mathbb{Z}_{2^L-1}$ symmetry. Furthermore, we find that the excitation gap above the ground state manifold does not decrease with the system size and is roughly close to the mean-field prediction, revealing the gapped nature of the breakdown condensate.

We have discussed in the main text that the excitation spectrum under open boundary conditions contains a zero-energy edge mode and several gapped bulk modes. As the edge mode is not compatible with periodic boundary conditions, we only observe plane-wave-like bulk modes here, which have a finite gap above the massively degenerate ground state manifold.  Given that the discrete symmetry asymptotically becomes a continuous exponential symmetry in the thermodynamic limit, the gaped bulk excitations are expected to hold under both periodic and open boundary conditions.

 \begin{figure}
    \centering
    \includegraphics[width=10cm]{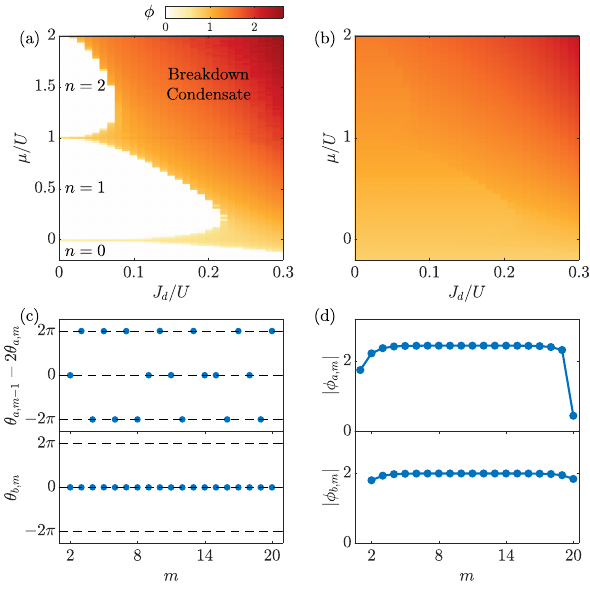}
    \caption{(a) The mean-field phase diagram of $\phi_a=\frac{1}{L}\sum_{m=1}^L|\braket{\hat a_m}|$. (b)  The mean-field phase diagram of $\phi_b=\frac{1}{L-1}\sum_{m=2}^L|\braket{\hat b_m}|$. (c) The phase angles $\theta_{a,m}=\arg\braket{\hat a_m}$ and $\theta_{b,m}=\arg\braket{\hat b_m}$ for $\mu/U=1.5,\ J_d/U=0.3$. (d) The field amplitude $|\phi_{a,m}|=|\braket{\hat a_m}|$  and $|\phi_{b,m}|=|\braket{\hat b_m}|$ for  $\mu/U=1.5,\ J_d/U=0.3$. In the variational calculation, the particle number cutoff is $N_{\max}=40$. Other parameters are $U=1, F=-0.5$.  }
    \label{sfig:practical}
\end{figure}

\section{The mean-field phase diagram for a practical model}
The main text introduces a practical model that can potentially realize exponential U(1) symmetry in state-of-the-art cold atom experiments. This section aims to provide a detailed mean-field calculation that supports the argument in the main text. 

Since the bosonic quantum breakdown Hubbard model discussed in the main text does not conserve the total boson number $\hat{N}_a = \sum_{m=1}^N \hat{a}_m^\dagger \hat{a}_m$, we introduce ancilla bosons $\hat{b}_m^\dagger$ at each site to serve as a particle reservoir. As mentioned in the main text, we consider the following Hamiltonian:
\begin{equation}
\begin{aligned}
        H=&-\sum_{m=1}^{L-1}[J_d\hat a_{m}(\hat a_{m+1}^{\dagger })^2\hat b_{m+1}+\text{h.c.}]+\sum_{m=1}^L[-\mu \hat a_m^\dagger \hat a_m+\frac{U}{2} \hat a_m^\dagger \hat a_m^\dagger \hat a_m \hat a_m]\\
        &+\sum_{m=2}^{L}[-\mu \hat b_m^\dagger \hat b_m+\frac{U}{2} \hat b_m^\dagger \hat b_m^\dagger \hat b_m \hat b_m+F(\hat b_{m}+\hat b_m^\dagger)].
\end{aligned}
\label{seq:pratical_H}
\end{equation}
In this Hamiltonian, the coupling $J_d$ in the first term of Eq.~\eqref{seq:pratical_H} mediates the interaction between system and ancilla bosons while preserving the total particle number $\hat{N}_a + \hat{N}_b$, where $\hat{N}_b = \sum_{m=2}^L \hat{b}_m^\dagger \hat{b}_m$. We emphasize that this interaction is experimentally realizable in cold-atom systems, sharing the same structure as dipole-conserving bosonic interactions. The second term in Eq.~\eqref{seq:pratical_H} describes the on-site chemical potential and Hubbard interaction for the system bosons, while the last term governs the ancilla bosons with an additional driving term $F$. Under certain conditions, the external driving frequency can be absorbed into the chemical potential via the rotating-wave approximation. Therefore, we simply restrict our analysis to zero-frequency driving. For the sake of simplicity, we further assume that both species share the same chemical potential and the same Hubbard interaction strength.

In the absence of driving ($F = 0$), the Hamiltonian conserves both the total particle number $\hat{N}_a + \hat{N}_b$ and the exponential charge $\hat{Q}_a = \sum_{m=1}^L 2^{L-m} \hat{a}_m^\dagger \hat{a}_m$. Introducing finite driving ($F \neq 0$) breaks the conventional U(1) symmetry associated with $\hat{N}_a + \hat{N}_b$ while preserving the exponential U(1) symmetry of $\hat{Q}_a$. Intuitively, the drive enables ancilla bosons to function as a particle source for the breakdown interaction, inducing a low-energy state with $\langle \hat{b}_m \rangle = \phi_{b,m} \neq 0$. Consequently, the effective dynamics in the system is governed by the $\gamma=0$ bosonic quantum breakdown Hubbard model in Eq.~(1) of the main text, with the modified coupling $J \rightarrow J_d\phi_b$, where we assume a uniform bulk configuration $\phi_{b,m} = \phi_b$ for all sites $m$.

The bosonic Hamiltonian in Eq. \eqref{seq:pratical_H} under open boundary conditions can also be investigated by the variational Gutzwiller mean-field method introduced in Sec.\ref{apsec:meanfield}.  The mean-field results are presented in Fig. \ref{sfig:practical}. As shown in Fig. \ref{sfig:practical}(b), in the whole parameter region, $ \phi_b=\frac{1}{L-1}\sum_{m=2}^L|\braket{\hat b_m}|$ is nonzero, as expected by the existence of external drives. Meanwhile, Figs. \ref{sfig:practical} (c) and (d) indicate that the phase angle and amplitude of $\phi_{b,m}=\braket{\hat b_m}$ exhibit a uniform pattern in the bulk, supporting our assumption in the last paragraph.  

Remarkably, $ \phi_a=\frac{1}{L}\sum_{m=2}^L|\braket{\hat a_m}|$ in Fig. \ref{sfig:practical}(a) demonstrates a similar mean-field phase diagram as Fig. 1(a) in the main text \footnote{There is also a discontinuous jump of \(\phi_b\) at the phase boundary determined by $\phi_a$, although small and less obvious in the colormap of Fig.~\ref{sfig:practical}(b). This is because the nonzero \(\phi_a\) value in the breakdown condensate phase effectively enhances the driving strength of the ancilla bosons via the \(J_d\) interaction. }. It shows a phase transition between several Mott insulating lobes and a breakdown condensate. Furthermore, in the breakdown condensate region where $\phi_{a,m}=\braket{\hat a_m}\ne0$, Fig. \ref{sfig:practical}(c) demonstrates that the phase angles $\theta_{a,m}=\arg\phi_{a,m}$ satisfy $\theta_{a,m-1}-2\theta_{a,m}=0 \mod 2\pi$ . These results indicate the spontaneous breaking of exponential U(1) symmetry for the Hamiltonian Eq. \eqref{seq:pratical_H}. 

Given that the model in Eq. \eqref{seq:pratical_H} is experimentally accessible, we expect that the spontaneous breaking of exponential U(1) symmetry can be observed  in state-of-the-art cold-atom platforms.

 \bibliography{boson_ref}